# Historical Context and Outlook of Quantum Hall Research for the Redefined SI

**Albert F. Rigosi**

Physical Measurement Laboratory, National Institute of Standards and Technology, Gaithersburg, MD 20899, United States of America

## 1. Basics of the Quantum Hall Effect

*1.1 Basics of the Quantum Hall Effect*

To fully appreciate the impacts that the discovery of the quantum Hall effect (QHE) had on electrical metrology, it may benefit the reader to cultivate a general understanding of the phenomenon [1, 2]. For the purposes of this handbook, a basic overview will be given. Two-dimensional (2D) electron systems can exhibit many interesting quantum phenomena [3]. The QHE may be exhibited by a 2D electron system when placed under a strong magnetic field perpendicular to the plane of the system. These conditions allow for Landau quantization, or the discretization of available energies of the electrons affected by the magnetic field. These energy values, determined by solving the Schrödinger equation, are known as Landau levels. During the course of a QHE measurement, one defines the Hall resistance $R_{xy}$ as the measured voltage, perpendicular to the direction of the applied current, divided by that same current. The characteristic longitudinal resistivity $\rho_{xx}$ goes to zero as $R_{xy}$ approaches a quantized value (nominally called a plateau).

Since the discovery of the QHE at the start of the 1980s, the electrical metrology community sought to implement a resistance standard based on the following relation: $R_H = h/\nu e^2$, where $\nu$ is an integer. Multiple quantized resistance steps were observable and while in the quantum Hall regime, an electron system could be measured electrically to yield a resistance that is expressible in terms of two fundamental constants. When the redefinition of the SI occurred in 2019, the constants $h$ and $e$, or the Planck constant and elementary charge, respectively, adopted a globally agreed-upon value that drastically changed how the electrical units of the ohm, volt, and ampere are defined. For instance, the von Klitzing constant changed from its conventional value set in 1990 ($R_{K-90}$ = 25 812.807 Ω) to the value of $h/e^2$ ($R_K$ = 25 812.807 459 304 5 Ω).

Before the QHE, resistance metrology would employ standard resistors made from copper-manganese-nickel and similar alloys. Since artifact standards exhibited a time-dependent variation of their nominal resistance [4], switching to a robust definition like the ratio of fundamental constants became the central focus of the field.

## 2. Predecessors for Quantum Hall Standards

Quantum Hall standards began their journey shortly after the QHE was discovered. Some of the earliest devices were based on silicon metal–oxide–semiconductor field-effect transistors (MOSFETs). For the case of these Si systems, the electric field needed to confine electrons to two dimensions was generated by a voltage gate and was separated from the surface of the semiconductor by an insulating oxide layer. In work performed by Hartland *et al.*, a cryogenic current-comparator (CCC) bridge with a sensitive 1:1- or 2:1-ratio was used to compare two quantized Hall resistances, one from the MOSFET, and the other from a gallium arsenide (GaAs) heterostructure [5]. More details on the CCC will be provided in a later section, so it should temporarily be treated as equipment that enables sensitive comparisons between two resistance outputs. Both devices were operated at the same temperature and magnetic field. The CCC had four 10000-turn windings surrounded by a superconducting shield. The magnetic flux was sensed by an 8-turn niobium coil connected to the input coil of a commercial 20 MHz RF superconducting quantum interference device (SQUID). The critical components of the CCC were immersed in liquid helium at 4.2 K and was shielded from external magnetic fields by using lead and low-temperature Mumetal [5].

The work focuses on comparing the quantized Hall resistance (QHR) of the GaAs/AlGaAs heterostructure (measured at the $\nu = 2$ plateau, or $h/2e^2$, which is approximately 12.9 k$\Omega$) to that of the silicon MOSFET device (measured at the $\nu = 4$ plateau, or $h/4e^2$, about 6.5 k$\Omega$). The deviations are represented as $\Delta_{24}$ and the ratio of the windings allows two resistances to be measured while considering the multiplicative factor of 2 that differs between them. These results are summarized in Ref. [5] and suggest that the two QHRs agree within $2[1-0.22(3.5) \times 10^{-10}]$. Given the novelty of the QHE at the time, this type of experiment was still supporting the notion of representing the plateaus as fundamental constants.

One of the largest disadvantages of MOSFETs was the high magnetic field requirement during an experiment. Higher magnetic fields typically yielded a wider resistance plateau. As seen in [5], the Si MOSFET device required 13 T to access its $\nu = 4$ plateau, and currents as low as 10 μA would cause the QHE to break down. These devices did not stay long since GaAs-based devices had also quickly demonstrated a more optimal ease-of-use [6, 7]. In GaAs-based devices, a 2D layer of electrons forms when an electric field forces electrons to the interface between two semiconductor layers (the other, in this case, being AlGaAs). For many devices with this type of interface, the layers are grown via molecular beam epitaxy [8-12]. Similar heterostructures have been developed in InGaAs/InP, which is obtained via metal-organic chemical vapor deposition [13].

These heterostructure devices have been grown with excellent homogeneity and exhibited high mobilities (on the order of 200 000 cm$^2$V$^{-1}$s$^{-1}$) at 4 K. During the 1980s, metrologists utilized these high-quality devices to confirm the universality of the von Klitzing constant $R_K$. Various national metrology institutes (NMIs) had improved the uncertainty of $R_K$ over the course of many experiments. The success of this material in providing highly precise access to quantized resistances led to its global adoption for representing the ohm around 1990 via the approval of $R_{K-90}$ by the Consultative Committee for Electricity (CCE) [14]. And for nearly twenty years, semiconductor technologies continued to improve, allowing both GaAs-based resistance standards and cryogenic measurement methods to become improved in themselves [15, 16].

An example of the use of GaAs-based devices for metrology can be seen in Cage *et al.* [6], where the authors looked to formally adopt GaAs as a standard used to maintain a laboratory unit of resistance. The work demonstrated the universality of the QHE and showed the viability of the device as a means to calibrate artifact standards. The devices were grown by molecular beam epitaxy and had dimensions as shown in the inset of Fig. 1 in Ref. [6]. The magnetic field sweep data for the Hall and longitudinal voltages are also shown there. The second part of the experiment involved calibrating a set of 6453.2 $\Omega$ resistors. This calibration required the use of two 6453.2 $\Omega$ resistors and 100 $\Omega$ series-parallel Hamon resistor networks [17]. Comparisons were done using a direct current comparator (DCC) resistance bridge (more information on the DCC will be provided in a later section) [6].

## 3 Expansion of QHR Device Capabilities

QHR devices soon became the norm in the electrical metrology community, with many of the NMI efforts implementing the new standards based on GaAs devices [18-22]. With the part-per-million changes that occurred with many NMI standard resistors from the 1990 redefinition, better agreement was obtained (by an order of magnitude) between the various worldwide resistance intercomparisons. The next natural step for metrologists was to examine whether or not these QHR devices could accommodate other values of resistance so that the calibration chain could be shortened. Such outputs may be accomplished by constructing quantum Hall array resistance standards (QHARS) [23-26].

For instance, in Oe *et al.*, a 10 kΩ QHARS device was designed and consisted of 16 Hall bars, a drastic improvement from previous work [23]. The nominal value of the device was measured to have a deviation of about 34 nΩ/Ω from exactly 10 kΩ (with the data based on $R_{K-90}$). The design and final device can be seen in Fig. 2 of Ref. [23]. The device was measured by what are now conventional means; *i.e.*, using a CCC to compare the device against an artifact or other QHR device. In this case, a 100 Ω standard resistor was used to verify that the array device agreed with its nominal value to within approximately one part in $10^8$. The work also proposed new combinations of Hall bars such that the array output could be customized for any of the decade values between 100 Ω and 1 MΩ [23].

In addition to the benefits gained from expanding GaAs-based devices further into the world of metrology using direct current (DC), expansion was also explored in the realm of alternating current (AC). Various NMIs had already begun standardizing impedance by using the QHE in an effort to replace the calculable capacitor, which is a difficult apparatus to construct [27-31]. As shown in Bykov *et al.*, linear AC resistances of the 2D electron system in GaAs-based devices was studied between excitations of 10 KHz and 20 GHz [29]. It was found that, while in the DC regime, the longitudinal resistances oscillated with current and externally applied magnetic field, much like what may be expected of other GaAs-based devices. However, when the applied current was AC, a new set of oscillations dependent on the magnetic field was found. This is relevant for impedance metrology because having a zero-resistance longitudinal resistance is one mark of a well-quantized device, and having AC-dependent oscillations would undoubtedly increase the uncertainty associated with the QHR. Challenges in adopting QHR technology still exist today in this branch of electrical metrology.

## 4 Limitations to the Modern QHR Technology

Given the historical context of resistance standards, and the expansive field that is involved with the modern day, graphene-based QHR standards [32-78], it would benefit those in the field to make an accurate assessment of which material properties or exhibitions are most influential for future metrological applications. To accomplish this goal, we must understand the limits of the current technology, at least to some rudimentary extent. For instance, it is not quite known with substantial certainty how high a current could be applied to a single EG hall bar before the QHE is guaranteed to break down. Additionally, we have not yet reached the upper bound on operation temperature that could be used for EG-based devices. Obviously, there are few enough limitations on single Hall bar EG-based QHR devices that the work thus far to establish this new global resistance standard has been and will likely remain virtually unimpeded.

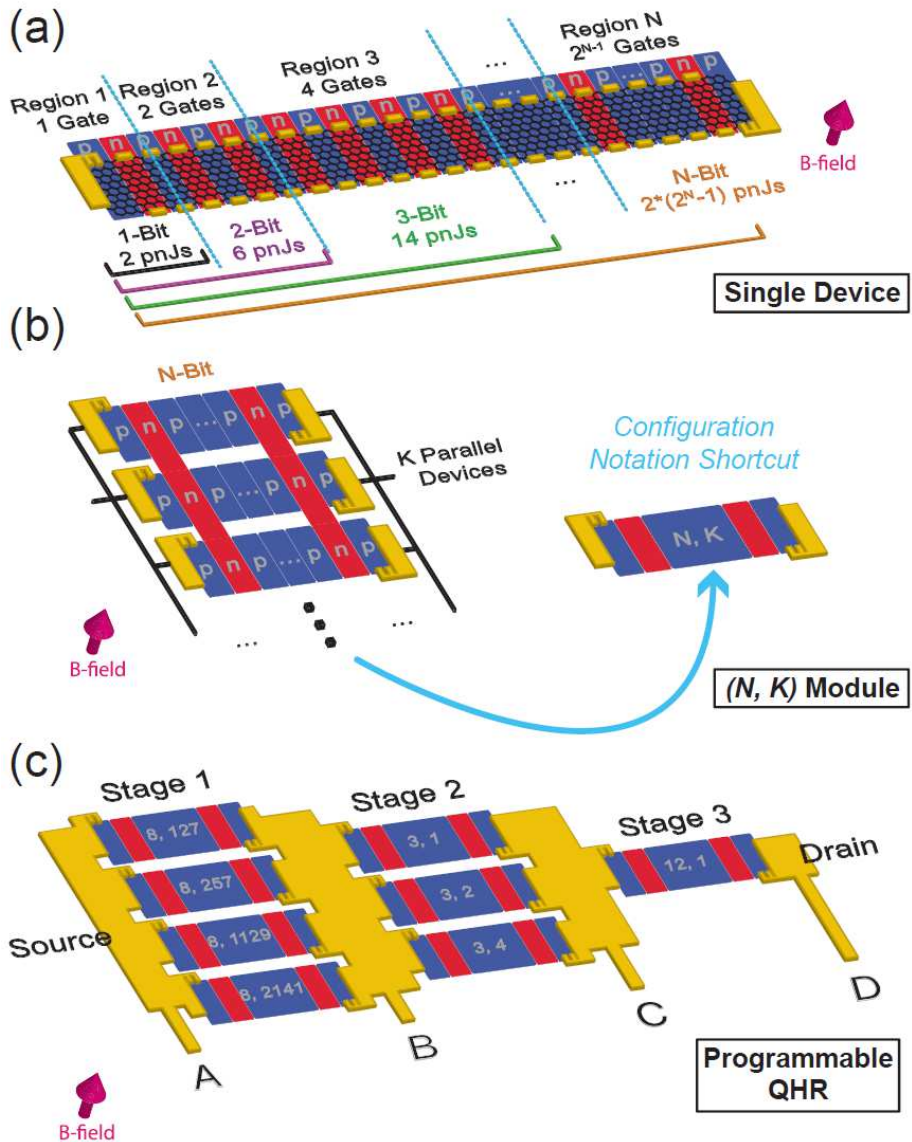

Figure 1. The proposed device illustrated represents a programmable QHR device for scalable standards. (a) An *N-bit* device is illustrated showing how each region is defined and the maximum number of *pn*Js that can be used. (b) This device, when connected in parallel with *K* copies of itself, becomes the foundation of the $(N, K)$ *module*. Each region has a set of gates that extend to all *K* parallel branches. (c) The proposed device is illustrated and composed of eight $(N, K)$ *modules*, four of which are in parallel in stage 1, three of which are parallel in stage 2, and a lone *module* in stage 3. All three stages are connected in series and all connections and contacts are proposed to be superconducting metal to eliminate the contact resistance to the greatest possible extent. The *modules* in stage 1 are *8-bit* devices with more than 100 parallel copies per *module*, whereas the *modules* in stage 2 are *3-bit* devices with four or fewer parallel copies per *module*. Stage 3 is a single *12-bit* device with no additional parallel branches. These numbers for $(N, K)$ are required should one wish to reproduce the values in Table 1. Ref. [62] is an open access article distributed under the terms of the Creative Commons CC BY license, which permits unrestricted use, distribution, and reproduction in any medium.

There is much research to be conducted in the case of newer technologies that build on the single Hall bar design despite already having gained some benefit from their development. For instance, the *pn*J devices have the widest access to different fractional or integer multiples of $R_H$ (or $R_K/2$) as output resistances, especially if source and drain currents are allowed to occupy more than two total terminals. If one defines $q$ as a coefficient of $R_H$, then the following relation may be used [79]:

$$q_{M-1}(n_{M-1}) = \frac{q_{M-2}(n_{M-1} + 1)}{n_{M-1} + \frac{q_{M-2}}{q_{M-1}^{(0)}}}$$

(1)

In equation (1), $M$ is the number of terminals in the *pn*J device, $n$ is the number of junctions between the outermost terminal and its nearest neighbor, and $q_{M-1}^{(0)}$ refers to a default value the device outputs when the configuration in question is modified such that its outermost terminal moves to share the same region as its nearest neighbor [79]. The key takeaway with this algorithm (equation 1) is that an incredibly vast set of available resistances becomes hypothetically possible by simple reverse engineering. The algorithm assumes that the Hall bar is of conventional linearity. That is, each $p$ region is adjacent to two other $n$ regions unless it is an endpoint. The same would hold true when $p$ and $n$ are swapped. The equation breaks down when the *pn*J device geometry changes to that of a checkerboard grid or Corbino-type geometry. In all of these cases, the available resistances in this parameter space are vastly abundant and will obviously not be a limiting factor for this species of device. Instead, limitations may stem from imperfections in the device fabrication, an almost inevitable manifestation as the device complexity increases [65].

In the limit of the purely hypothetical, should the resistance metrology community wish to scale to decade values only, as per the existing infrastructure, then a programmable resistance standard may be able to provide many decades of quantized resistance output by following the designs proposed by Hu *et al.* [62]. The proposed programmable QHR device is illustrated in Fig. 1, with each subfigure defining a small component of a total device. When programmed in a particular way, this single device can output all decades between 100 Ω and 100 MΩ, as summarized by Table 1.

| Resistance | Stage 1 | Stage 2 | Stage 3 | Voltage probes used | Deviation from decade value |
|---|---|---|---|---|---|
| **100 Ω** | 00010010 | None | None | A, B | 0.714 µΩ/Ω |
|  | 00011001 |  |  |  |  |
|  | 00001001 |  |  |  |  |
|  | 00010001 |  |  |  |  |
| **1 kΩ** | 10001001 | None | None | A, B | 0.108 µΩ/Ω |
|  | 10001011 |  |  |  |  |
|  | 10000111 |  |  |  |  |
|  | 10010001 |  |  |  |  |
| **10 kΩ** | 00110001 | 011 | None | A, C | 14.8 nΩ/Ω |

|  | 00100111 | 010 | | | |
|  | 00110011 | 010 | | | |
|  | 00111100 | | | | |
| **100 kΩ** | 00111101 | 010 | 000000000011 | A, D | 0.043 nΩ/Ω |
|  | 00111101 | 010 | | | |
|  | 00111100 | 010 | | | |
|  | 00111110 | | | | |
| **1 MΩ** | 00101011 | 010 | 000000100110 | A, D | 0.0243 nΩ/Ω |
|  | 00100001 | 001 | | | |
|  | 00110101 | 001 | | | |
|  | 00110000 | | | | |
| **10 MΩ** | 01011110 | 110 | 000110000010 | A, D | 0.346 pΩ/Ω |
|  | 01011001 | 101 | | | |
|  | 01010011 | 110 | | | |
|  | 01001100 | | | | |
| **100 MΩ** | 01110111 | 100 | 111100100001 | A, D | 1.21 nΩ/Ω |
|  | 10001001 | 011 | | | |
|  | 01101101 | 011 | | | |
|  | 01100111 | | | | |

Table 1. This shows all possible resistance decades achievable with the proposed programmable QHR device in Fig. 1. The values listed in this table are described in more detail in Ref. [62] and can achieve very accurate values of seven decades of resistance. Each module in each stage is assigned a binary string. As long as the exact configuration is used, the measured voltage between the two probes will measure a near-exact decade value, to within a deviation defined in the rightmost column. Ref. [62] is an open access article distributed under the terms of the Creative Commons CC BY license, which permits unrestricted use, distribution, and reproduction in any medium.

With all *pn*J devices, better gating techniques are warranted. In the case of top gating, which is the basis for forming the device measured by Hu *et al*. [62], fabrication is limited by the size of the exfoliated *h*-BN flake typically used as a high-quality dielectric spacer. For bottom gating of EG on SiC, as seen with ion implantation [80], there has not yet been a demonstration of metrologically viable devices, though there may still be potential for perfecting this technique. Since gating is likely to be the largest limiting factor for *pn*Js, one must instead turn to array technology. Although QHARS devices can theoretically replicate *pn*Js, the necessity of an interconnection will ultimately result in a smaller available parameter space. Nonetheless, the output quantized resistance values offered by this technology may still be sufficiently plentiful for future applications in electrical metrology.

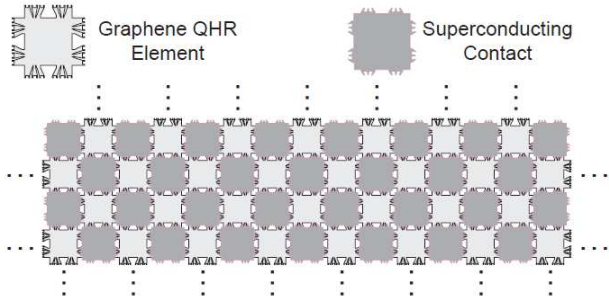
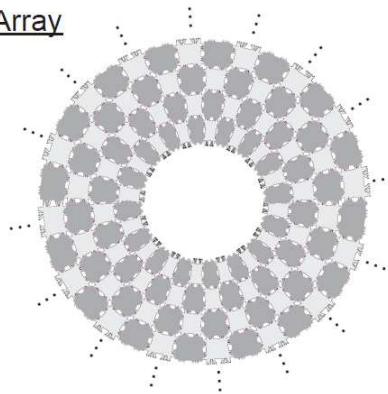
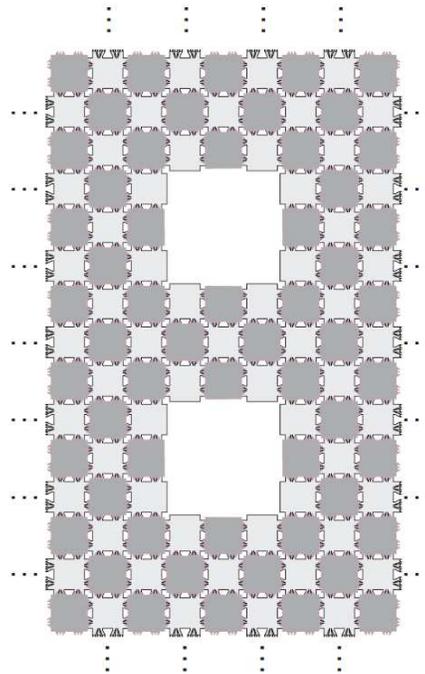

Figure 2. Hypothetical array designs of varying topological genera. Determining the predicted value of output quantized resistance between any pair of contacts can be done with various modeling techniques done for similar systems. (a) The use of superconducting contacts enables the design of larger and more complex arrays, provided the EG QHR elements are small enough. For genus 0, grid arrays can take on user-defined dimensions. (b) Corbino-type geometries could also be implemented, and these are just examples of genus 1 topologies. (c) A final example of array type comes from those that have a genus 2 topology. Even more customized designs are possible and not well-explored.

When it comes to arrays, there are several types that can take shape for a potential QHR device. There are the conventional parallel or series devices, which are the subject of recent works [61, 67-70]. As one departs from this simpler design, the number of potential quantized resistances that become available rapidly increases. Designs with varying topological genera, as shown in Fig. 2, along with the predictive power of simulations like LTspice, Kwant, or traditional tight-binding Hamiltonians, allows the designer to customize devices accordingly. Depending on the genus and final layout, QHARS devices can still be metrologically verified by means of measuring a specific configuration and its mirror symmetric counterparts. In that sense, future QHARS devices whose longitudinal resistances cannot be checked must either be compared with a duplicate of itself, or must have appropriate symmetries such that a same-device comparison can be made. For instance, a square array could be the subject of a two-terminal measurement using same-sided corners. This configuration would have a four-fold symmetry which can be measured and compared to verify device functionality. All types of arrays would be thus limited by the EG growth area, which at present has been optimized by the use of a polymer-assisted sublimation growth technique [51]. The total growth area may be the most demanding limiting factor for this species of device. Nonetheless, one can hope that the latter technique, and any similar technique to be developed, will enable homogeneous growth on the wafer scale such that the whole EG area retains metrological quality.

**5. The Quantum Anomalous Hall Effect**

Regardless of EG device size, the magnetic field requirement will always be a limit. This is inherently tied to the band structure of graphene. In addition to this limitation, there are at least two others that prevent SI-traceable quantum electrical standards beyond the ohm from being user-friendly and more widely disseminated, which at present, confines their global accessibility to just NMIs. These other limitations include the sub-nA currents obtained from single electron transistors (in the case of the quantum ampere) and the Josephson voltage standard's aversion to magnetism (which, in this approach, must be housed with a QHR device to create a compact current standard). Ongoing research on topological material systems has the potential to solve these major compatibility problems.

The physical phenomenon underpinning this research is the quantum anomalous Hall effect (QAHE). This effect yields a quantized conductance in magnetically ordered materials at zero applied magnetic field. The QAHE is a manifestation of a materials' topologically nontrivial electronic structure. The QAHE, along with the Josephson effect and QHE, is a rare example of a macroscopic quantum phenomenon. There are several types of materials that exhibit the QAHE, with many being classified within the following categories: magnetically doped topological insulators, intrinsic magnetic topological insulators, and twisted van der Waals layered systems. An advantage to using TIs is that one many operate them at zero-field for measurements, as has been shown in some recent work [81, 82].

Fox *et al*. explored the potential of the QAHE in a magnetic TI thin film for metrological applications. Using a CCC system, they measured the quantization of the Hall resistance to within one part per million and, at lower current bias, measured the longitudinal resistivity under 10 m$\Omega$ at zero magnetic field [81]. An example of the data they acquired is in Fig. 2 of Ref. [81]. When the current density was increased past a critical value, a breakdown of the quantized state was induced, and this effect was attributed to electron heating in bulk current flow. Their work furthered the understanding of TIs by gaining a comprehension of the mechanism during the prebreakdown regime. The results had included evidence for bulk dissipation, including thermal activation and possible variable-range hopping. A concurrently reported work by Gotz *et al*. also looked to present a metrologically comprehensive measurement of a TI system (V-doped $(Bi,Sb)_2Te_3$) in zero magnetic field [82]. When they measured the deviation of the quantized anomalous Hall resistance from $R_K$, they determined a value of $0.176 \pm 0.25$ $\mu\Omega/\Omega$. An example of their data is shown in Fig. 5 of Ref. [82]. The steps both works made are vital to our eventual realization of a zero-field quantum resistance standard.

One of the remaining major limitations, besides finding a TI material system with a large band gap, will be to lift the stringent temperature requirements, which are currently in the 10 mK to 100 mK range. Fijalkowski *et al*. show, through a careful analysis of non-local voltages in devices having a Corbino geometry, that the chiral edge channels closely tied to the observation of the QAHE continue to exist without applied magnetic field up to the Curie temperature (20 K) of bulk ferromagnetism of their TI system. Furthermore, it was found that thermally activated bulk conductance was responsible for quantization breakdown [83]. The results give hope that one may utilize the topological protection of these edge channels for developing a QHR, as has been demonstrated most recently by Okazaki *et al*. [84]. In their work, they demonstrate a precision of 10 n$\Omega/\Omega$ of the Hall resistance quantization in the QAHE. They directly compared both the QAHE and QHE from a conventional device to confirm their observations of an accurate QAHE. Given this very recent development, more efforts are expected to follow to verify the viability of TIs as a primary standard for resistance. In the ideal case scenario, TI-based QHR devices will make disseminating the ohm more economical and portable, and will, more importantly, serve as a basis for a compact quantum ampere.

## 6. Outlook on the Field of Resistance Metrology

As the global implementation of new technologies continues to progress, we hope to see a more universal accessibility to the quantum SI. This chapter has given historical context for the role of the QHE in metrology, including a basic overview of the QHE, preceding device technologies, and how those devices were expanded in their capabilities. Next, the graphene era was summarized in terms of how the new 2D material performed compared with GaAs-based QHR devices, how the world began to implement it as a resistance standard, and how the corresponding measurement infrastructure has adapted to the new standard. In the third section, emerging technologies based on graphene were introduced to give a brief overview of the possible expansion of QHR device capabilities. These ideas and research avenues include *pn*J devices, QHARS devices, and experimental components of AC metrology and the quantum ampere. The chapter then concludes by discussing the possible limitations of graphene-based technology for resistance metrology and looks to explore topological insulators as one potential candidate to, at the very least, supplement graphene-based QHR devices for resistance and electrical current metrology.

It has become evident throughout the last few decades that the quantum Hall effect, as exhibited by our modern 2D systems both with and without magnetic fields, has the marvelous potential to unify the components of Ohm's law. That is, the QHE can bring together all three electrical quantities provided that the magnetic field requirement is small or irrelevant to the apparatus in use. Developing and deploying a system with several traceability capabilities will undoubtedly improve the status of electrical metrology worldwide. Throughout all the coming advancements, it will be important to remember that these milestones should keep us motivated to continue learning how to better enrich society with the quantum Hall effect:

> "It is characteristic of fundamental discoveries, of great achievements of intellect,
> that they retain an undiminished power upon the imagination of the thinker."
> – Nikola Tesla, 1891, New York City, New York

**Acknowledgements**

The authors wish to acknowledge S. Mhatre, A. Levy, G. Fitzpatrick, and E. Benck for their efforts and assistance during the internal review process at NIST. Commercial equipment, instruments, and materials are identified in this paper in order to specify the experimental procedure adequately. Such identification is not intended to imply recommendation or endorsement by the National Institute of Standards and Technology or the United States government, nor is it intended to imply that the materials or equipment identified are necessarily the best available for the purpose.

**References**

[1] v Klitzing K, Ebert G. Application of the quantum Hall effect in metrology. Metrologia. 1985;21(1):11.
[2] Von Klitzing K, Dorda G and Pepper M 1980 *Phys. Rev. Lett.* **45** 494
[3] Hill HM, et al. 2019 *Phys. Rev. B*. **99** 174110
[4] Witt TJ 1998 Rev. Sci. Instrum. **69** 2823-43
[5] Hartland A, Jones K, Williams J M, Gallagher B L and Galloway T 1991 *Phys. Rev. Lett.* **66** 969-73
[6] Cage ME, Dziuba RF, Field BF 1985 *IEEE Trans. Instrum. Meas.* **2,** 301-3.

[7] Hartland A 1992 *Metrologia* **29** 175
[8] Tsui D C and Gossard A C 1981 *Appl. Phys. Lett.* **38** 550
[9] Van Der Wel W, Harmans KJ, Kaarls R, Mooij JE 1985 *IEEE Trans. Instrum. Meas*. **2** 314-6
[10] Hartland A, Jones R G, Kibble B P and Legg D J 1987 *IEEE Trans. Instrum. Meas.* **IM-36** 208
[11] Bliek L, Braun E and Melchert F 1983 *Metrologia* **19**, 83
[12] Hartland A, Davis GJ and Wood DR 1985 *IEEE Trans. Instrum. Meas.* **IM-34** 309
[13] Delahaye F, Dominguez D, Alexandre F, André J P, Hirtz J P and Razeghi M 1986 *Metrologia* **22** 103-10
[14] Taylor BN 1990 *IEEE Trans. Instrum. Meas*. **39** 2-5
[15] Jeckelmann B, Jeanneret B and Inglis D. 1997 *Phys. Rev. B* **55** 13124
[16] Williams JM IET 2011 *Sci. Meas. Technol.* **5** 211-24
[17] Hamon BV 1954 *J. Sci. Instrum.* **31** 450-453
[18] Small GW, Ricketts BW, and Coogan PC 1989 *IEEE Trans. Instrum. Meas.* **38** 245
[19] Cage ME, Dziuba RF, Elmquist RE, Field BF, Jones GR, Olsen PT, Phillips WD, Shields JQ, Steiner RL, Taylor BN, and Williams ER 1989 *IEEE Trans. Instrum. Meas.* **38** 284
[20] Shields JQ and Dziuba RF 1989 *IEEE Trans. Instrum. Meas.* **38** 249
[21] Delahaye F and Jeckelmann B 2003 *Metrologia* **40** 217
[22] Jeckelmann B and Jeanneret B 2001 *Rep. Prog. Phys.* **64** 1603
[23] Oe T, Matsuhiro K, Itatani T, Gorwadkar S, Kiryu S, Kaneko NH 2013 *IEEE Trans. Instrum. Meas.* **62** 1755-9
[24] Poirier W, Bounouh A, Piquemal F and André JP 2004 *Metrologia* **41** 285
[25] Ortolano M, Abrate M and Callegaro L 2014 *Metrologia* **52** 31
[26] Konemann J, Ahlers FJ, Pesel E, Pierz K and Schumacher HW 2011 *IEEE Trans. Instrum. Meas.* **60** 2512-6
[27] Ahlers FJ, Jeanneret B, Overney F, Schurr J, Wood BM 2009 *Metrologia* **46** R1
[28] Cabiati F, Callegaro L, Cassiago C, D'Elia V, Reedtz GM. 1999 *IEEE Trans. Instrum. Meas.* **48** 314-8
[29] Bykov AA, Zhang JQ, Vitkalov S, Kalagin AK, Bakarov AK 2005 *Phys. Rev. B*. **72** 245307
[30] Hartland A, Kibble BP, Rodgers PJ, Bohacek J. 1995 *IEEE Trans. Instrum. Meas.* **44** 245-8
[31] Wood HM, Inglis AD, Côté M 1997 *IEEE Trans. Instrum. Meas* **46** 269-72
[32] Zhang Y, Tan YW, Stormer HL and Kim P 2005 Nature **438** 201
[33] Novoselov KS, Jiang Z, Zhang Y, Morozov SV, Stormer HL, Zeitler U, Maan JC, Boebinger GS, Kim P, Geim AK 2007 Science **315** 1379
[34] Novoselov KS, Geim AK, Morozov S, Jiang D, Katsnelson M, Grigorieva I, Dubonos S, Firsov AA 2005 Nature **438** 197
[35] De Heer WA, Berger C, Wu X, First PN, Conrad EH, Li X, Li T, Sprinkle M, Hass J, Sadowski ML, Potemski M 2007 *Solid State Commun*. **143** 92-100
[36] Jabakhanji B, Michon A, Consejo C, Desrat W, Portail M, Tiberj A, Paillet M, Zahab A, Cheynis F, Lafont F, Schopfer F. 2014 *Phys. Rev. B* **89** 085422
[37] Janssen TJ, Williams JM, Fletcher NE, Goebel R, Tzalenchuk A, Yakimova R, Lara-Avila S, Kubatkin S, Fal'ko VI 2012 *Metrologia* **49** 294
[38] Giesbers AJ, Rietveld G, Houtzager E, Zeitler U, Yang R, Novoselov KS, Geim AK, Maan JC 2008 *Appl. Phys. Lett.* **93** 222109–12
[39] Tzalenchuk A Lara-Avila S, Kalaboukhov A, Paolillo S, Syväjärvi M, Yakimova R, Kazakova O, Janssen TJ, Fal'Ko V, Kubatkin S 2010 *Nat. Nanotechnol*. **5** 186–9
[40] Lafont F, Ribeiro-Palau R, Kazazis D, Michon A, Couturaud O, Consejo C, Chassagne T, Zielinski M, Portail M, Jouault B, Schopfer F 2015 *Nat. Commun.* **6** 6806
[41] Janssen T J B M, Tzalenchuk A, Yakimova R, Kubatkin S, Lara-Avila S, Kopylov S, Fal'ko VI. 2011 *Phys. Rev. B* **83** 233402–6
[42] Oe T, Rigosi AF, Kruskopf M, Wu BY, Lee HY, Yang Y, Elmquist RE, Kaneko N, Jarrett DG 2019 *IEEE Trans. Instrum. Meas.* 2019 **69** 3103-3108
[43] Woszczyna M, Friedemann M, Götz M, Pesel E, Pierz K, Weimann T, Ahlers FJ 2012 *Appl. Phys. Lett.* **100** 164106
[44] Satrapinski A, Novikov S, Lebedeva N 2013 *Appl. Phys. Lett.* **103** 173509
[45] Rigosi AF, Panna AR, Payagala SU, Kruskopf M, Kraft ME, Jones GR, Wu BY, Lee HY, Yang Y, Hu J, Jarrett DG, Newell DB, and Elmquist RE 2019 *IEEE Trans. Instrum. Meas*. **68**, 1870-1878.
[46] Lara-Avila S, Moth-Poulsen K, Yakimova R, Bjørnholm T, Fal'ko V, Tzalenchuk A, Kubatkin S 2011 *Adv. Mater.* **23** 878-82
[47] Rigosi AF, Liu CI, Wu BY, Lee HY, Kruskopf M, Yang Y, Hill HM, Hu J, Bittle EG, Obrzut J, Walker AR 2018 *Microelectron. Eng*. **194** 51-5
[48] Riedl C, Coletti C, Starke U 2010 *J. Phys. D* **43** 374009
[49] Rigosi AF, Hill HM, Glavin NR, Pookpanratana SJ, Yang Y, Boosalis AG, Hu J, Rice A, Allerman AA, Nguyen NV, Hacker CA, Elmquist RE, Newell DB 2017 *2D Mater.* **5** 011011
[50] Janssen TJ, Rozhko S, Antonov I, Tzalenchuk A, Williams JM, Melhem Z, He H, Lara-Avila S, Kubatkin S, Yakimova R 2015 *2D Mater.* **2** 035015
[51] Kruskopf M, Pakdehi DM, Pierz K, Wundrack S, Stosch R, Dziomba T, Götz M, Baringhaus J, Aprojanz J, Tegenkamp C, Lidzba J. 2016 *2D Mater.* **3** 041002

[52] Hill HM, Rigosi AF, Chowdhury S, Yang Y, Nguyen NV, Tavazza F, Elmquist RE, Newell DB, Walker AR. 2017 *Phys. Rev. B* **96** 195437
[53] Rigosi AF, Elmquist RE 2019 *Semicond. Sci.Ttechnol*. 2019 **34** 093004
[54] MacMartin MP, Kusters NL 1996 *IEEE Trans. Instrum. Meas.* **15** 212-20
[55] Drung D, Götz M, Pesel E, Storm JH, Aßmann C, Peters M, Schurig T 2009 *Supercond. Sci. Technol.* **22** 114004
[56] Sullivan DB and Dziuba RF 1974 *Rev. Sci. Instrum*. **45** 517
[57] Grohmann K, Hahlbohm HD, Lübbig H, and Ramin H 1974 *Cryogenics* **14** 499
[58] Williams J M, Janssen T J B M, Rietveld G and Houtzager E 2010 *Metrologia* **47** 167–74
[59] Zhang N 2006 *Metrologia* **43** S276-S281
[60] Delahaye F 1993 *J. Appl. Phys*. **73** 7914-20
[61] Panna AR, Hu IF, Kruskopf M, Patel DK, Jarrett DG, Liu CI, Payagala SU, Saha D, Rigosi AF, Newell DB, Liang CT 2021 *Phys. Rev. B* **103** 075408
[62] Hu J, Rigosi AF, Kruskopf M, Yang Y, Wu BY, Tian J, Panna AR, Lee HY, Payagala SU, Jones GR, Kraft ME, Jarrett DG, Watanabe K, Takashi T, Elmquist RE, Newell DB 2018 *Sci. Rep.* **8** 15018
[63] Woszczyna M, Friedemann M, Dziomba T, Weimann T, Ahlers FJ 2011 *Appl. Phys. Lett.* **99** 022112
[64] Hu J, Rigosi AF, Lee JU, Lee HY, Yang Y, Liu CI, Elmquist RE, Newell DB 2018 *Phys. Rev B* **98** 045412
[65] Rigosi AF, Patel DK, Marzano M, Kruskopf M, Hill HM, Jin H, Hu J, Hight Walker AR, Ortolano M, Callegaro L, Liang CT, Newell DB 2019 *Carbon* **154** 230-237
[66] Momtaz ZS, Heun S, Biasiol G, Roddaro S 2020 *Phys. Rev. Appl.* **14** 024059
[67] Kruskopf M, Rigosi AF, Panna AR, Marzano M, Patel DK, Jin H, Newell DB, Elmquist RE 2019 *Metrologia* **56** 065002
[68] He H, Cedergren K, Shetty N, Lara-Avila S, Kubatkin S, Bergsten T, Eklund G 2021 *arXiv* preprint arXiv:2111.08280
[69] Kruskopf M, Rigosi AF, Panna AR, Patel DK, Jin H, Marzano M, Newell DB, Elmquist RE 2019 *IEEE Trans. Electron Devices* **66** 3973-3977
[70] Park J, Kim WS, Chae DH 2020 *Appl. Phys. Lett.* **116** 093102
[71] Clothier WK 1965 *Metrologia* **1** 36
[72] Cutkosky RD 1974 IEEE Trans. Instrum. Meas. **23** 305-9
[73] Kruskopf M, Bauer S, Pimsut Y, Chatterjee A, Patel DK, Rigosi AF, Elmquist RE, Pierz K, Pesel E, Götz M, Schurr J 2021 *IEEE Trans. Electron Devices* **68** 3672-3677
[74] Lüönd F, Kalmbach CC, Overney F, Schurr J, Jeanneret B, Müller A, Kruskopf M, Pierz K, Ahlers F 2017 *IEEE Trans. Instrum. Meas.* **66** 1459-66
[75] Giblin SP, Kataoka M, Fletcher JD, See P, Janssen TJ, Griffiths JP, Jones GA, Farrer I, Ritchie DA 2012 *Nat. Commun.* 2012 **3** 1-6
[76] Pekola JP, Saira OP, Maisi VF, Kemppinen A, Möttönen M, Pashkin YA, Averin DV 2013 *Rev. Mod. Phys.* **85** 1421
[77] Koppinen PJ, Stewart MD, Zimmerman NM 2012 *IEEE Trans. Electron Devices* **60** 78-83
[78] Brun-Picard J, Djordjevic S, Leprat D, Schopfer F, Poirier W 2016 *Phys. Rev. X* **6** 041051
[79] Rigosi AF, Marzano M, Levy A, Hill HM, Patel DK, Kruskopf M, Jin H, Elmquist RE, Newell DB 2020 *Phys. B: Condens. Matter* **582** 411971
[80] Waldmann D, Jobst J, Speck F, Seyller T, Krieger M, Weber HB 2011 *Nat. Mater.* **10** 357-60
[81] Fox EJ, Rosen IT, Yang Y, Jones GR, Elmquist RE, Kou X, Pan L, Wang KL, and Goldhaber-Gordon D 2018 *Phys. Rev. B* **98** 075145
[82] Götz M, Fijalkowski KM, Pesel E, Hartl M, Schreyeck S, Winnerlein M, Grauer S, Scherer H, Brunner K, Gould C, Ahlers FJ 2018 *Appl. Phys. Lett.* **112** 072102
[83] Fijalkowski KM, Liu N, Mandal P, Schreyeck S, Brunner K, Gould C, Molenkamp LW 2021 *Nat. Commun.* 2021 **12** 1-7
[84] Okazaki Y, Oe T, Kawamura M, Yoshimi R, Nakamura S, Takada S, Mogi M, Takahashi KS, Tsukazaki A, Kawasaki M, Tokura Y 2021 *Nat. Phys.* 13:1-5